\documentclass[12pt,aps,prd,preprint,tightenlines,superscriptaddress,showpacs,nofootinbib]
{revtex4-1}
\usepackage{epsfig}
\usepackage{amssymb,amsmath}
\usepackage{hyperref}
\usepackage{xfrac}
\usepackage{color}
\usepackage[utf8]{inputenc} 
\usepackage{feynmp-auto}

\newcommand{\bea}{\begin{eqnarray}}
\newcommand{\eea}{\end{eqnarray}}

\begin{document}
%
\vspace*{1.0cm}

\begin{center}
\baselineskip 20pt 
{\Large\bf 
Endothermic $Z^\prime$-Portal Dark Matter: \\
LZ-LHC Complementarity 
}
\vspace{1cm}

{\large
Nobuchika Okada$^{~a,}$\footnote{okadan@ua.edu}, 
and Digesh Raut$^{~b,}$\footnote{draut@smcm.edu}
}
\vspace{.5cm}

{\baselineskip 20pt \it
$^{a}$ Department of Physics and Astronomy, \\ The University of Alabama, Tuscaloosa, Alabama 35487, USA \\
$^{b}$ Physics Department, \\St. Mary's College of Maryland, St. Mary's City, Maryland 20686, USA
} 

\vspace{.5cm}

\vspace{1.5cm} {\bf Abstract}
\end{center}

Motivated by the $2.6\sigma$ high-energy recoil event recently reported by the LUX-ZEPLIN (LZ) Collaboration, we consider an endothermic $Z^\prime$-portal Majorana dark matter framework and discuss the complementarity between the LZ event and LHC searches for a $Z^\prime$ resonance. As a concrete realization, we consider a gauged U(1) ${B-L}$ extension of the Standard Model. The phenomenology of the framework is essentially controlled by two free parameters: the U(1) ${B-L}$ gauge coupling $g_{BL}$ and the $Z^\prime$ boson mass $m_{Z^\prime}$. For a fixed $m_{Z^\prime}$, the observed dark matter abundance requires the dark matter mass to be near the $Z^\prime$ resonance, $m_{\rm DM} \sim m_{Z^\prime}/2$, and sets a lower bound on $g_{BL}$. Complementarily, LHC searches for a $Z^\prime$ resonance set an upper bound on $g_{BL}$. The parameter space allowed by the dark matter abundance and LHC constraints can account for the recent LZ event. The synergy between future $Z^\prime$ resonance searches at the High-Luminosity LHC and the LZ experiment may provide a test of this framework.

\thispagestyle{empty}

\newpage

\addtocounter{page}{-1}


\section{Introduction}

Recently, the LUX-ZEPLIN (LZ) Collaboration reported a nuclear recoil event at
$E_R=248\pm23\,({\rm stat})\pm23\,({\rm sys})~{\rm keV}$, with a local
significance of $3.4\sigma$ and a global significance of
$2.6\sigma$~\cite{LZ:2026axp}.
Such a high-energy recoil is difficult to accommodate with conventional elastic
scattering of halo dark matter (DM) particles off nuclei and has stimulated
considerable interest in possible interpretations of this event.
A variety of scenarios have been proposed, including inelastic DM scattering,
momentum- or spin-dependent DM interactions, boosted DM, DM absorption, and
possible neutrino or nuclear origins~\cite{Lou:2026idn, Jeesun:2026vzo, Dent:2026bji, deLima:2026shq, Gu:2026vto, Baer:2026fpy, Liang:2026coz, Alhazmi:2026efz, Kannike:2026qyl, Aghaie:2026vsu, Fan:2026hzw, Chattaraj:2026fxn, Heikinheimo:2026kwp, Lee:2026zbr, Mahapatra:2026glu, An:2026pkc}.
Among these possibilities, endothermic DM scattering, in which a DM particle
transitions to a slightly heavier state, is particularly interesting since it
naturally favors high-energy nuclear recoils~\cite{Su:2026rwz, Fan:2026kxx, Freese:2026sga, Wu:2026nhi, Yin:2026jnn, Nomura:2026qyq, DiMauro:2026ldr, Visinelli:2026kgt, Yamashita:2026ump, Smirnov:2026aqk, Du:2026guj, McCabe:2026crm, Unwin:2026rdp, Lee:2026wof, Wang:2026ytg, Yang:2026wpb, Das:2026uyy, Okada:2026eol, Ahmed:2026qjg, Du:2026lpa, Bandyopadhyay:2026gjw, Borah:2026zwf, Bisal:2026khf, Yuan:2026djt, Zhu:2026dag, Asadi:2026iot, Lee:2026xxh, Lee:2026jxl, Langhoff:2026ujr, He:2026hqz, Qi:2026vyp, Kumar:2026lgi, Frolovsky:2026tvq, Okada:2026upm, Borah:2026ris, Das:2026buc, He:2026idw}.
Several particle-physics realizations of this scenario have already been investigated including test of the complementarity and its interpretation~\cite{Pospelov:2026ewn, Rodd:2026tyn, DiMauro:2026dqp, Kotlarski:2026pep, Cheung:2026byg, Bose:2026ndd, Chatterjee:2026scv, Nguyen:2026lui, Palmisano:2026kuj, Bose:2026szs}

The thermally averaged DM annihilation cross section required to reproduce the observed relic abundance through standard thermal freeze-out is typically of order $1~{\rm pb}$. 
On the other hand, the interaction strength required to account for the LZ event is strongly constrained by direct-detection observations. 
If the same interaction controls both DM annihilation and DM--nucleon scattering, simultaneously reproducing the observed relic abundance and the LZ event can therefore be nontrivial. 
In particular, an additional mechanism that enhances the DM annihilation rate 
without correspondingly enhancing the direct-detection rate may be required.

In this work, we present a simple $Z^\prime$-portal framework for endothermic
DM that is largely independent of the underlying gauge extension of the SM. 
We consider two Majorana fermions, $\psi_1$ and $\psi_2$,
interacting with the SM through a $Z^\prime$ portal, where $\psi_1$ is
identified as the DM candidate. The interactions relevant to our analysis are
given by \footnote{For the $Z^\prime$ portal inelastic scalar DM scenario, see Ref.~\cite{Okada:2019sbb}.}
\bea
{\cal L} \supset
i g_{12}\,\overline{\psi}_1\gamma^\mu\psi_2\,Z^\prime_\mu
+ g_q\,\bar q\gamma^\mu q\,Z^\prime_\mu .
\eea
For a small mass splitting,
$m_2-m_1={\cal O}(10^2)\,{\rm keV}$, the DM particle can undergo endothermic inelastic scattering off nuclei, $\psi_1 N\to\psi_2 N$, through $Z^\prime$ exchange, potentially accounting
for the high-energy nuclear-recoil event reported by the LZ Collaboration.
At the same time, the $Z^\prime$-mediated DM annihilation can be resonantly
enhanced for $m_1\simeq m_{Z^\prime}/2$, allowing the observed DM relic
abundance to be reproduced while maintaining the scattering rate relevant
to the LZ event. Since the same $Z^\prime$ mediator can be resonantly
produced at the LHC, collider searches provide a complementary probe of the
parameter space relevant to DM phenomenology. We show that the interplay
among the LZ event, the DM relic abundance, and $Z^\prime$ resonance searches
at the LHC provides a powerful test of this framework, with future searches
at the High-Luminosity LHC offering further opportunities to probe the
remaining parameter space.

The rest of this paper is organized as follows.
In Sec.~2, we introduce the endothermic $Z^\prime$-portal DM framework
and present its realization in the gauged $U(1)_{B-L}$ model.
In Sec.~3, we study the DM phenomenology, including the relic abundance
and inelastic DM--nucleon scattering relevant to the LZ event.
In Sec.~4, we discuss the LHC constraints on the $Z^\prime$ boson and
their complementarity with the DM phenomenology.
We summarize our results in Sec.~5.

\section{Endothermic $Z^\prime$-Portal Dark Matter}

\begin{table}[ht!]
\begin{center}
\centering
\begin{tabular}{c|cccc|c}
\hline
 & $SU(3)_c$ & $SU(2)_L$ & $U(1)_Y$ & $U(1)_{B-L}$ & $Z_2$ \\
\hline
$q_L^i$ & $\mathbf{3}$ & $\mathbf{2}$ & $\frac{1}{6}$ & $\frac{1}{3}$ & $+$ \\
$u_R^i$ & $\mathbf{3}$ & $\mathbf{1}$ & $\frac{2}{3}$ & $\frac{1}{3}$ & $+$ \\
$d_R^i$ & $\mathbf{3}$ & $\mathbf{1}$ & $-\frac{1}{3}$ & $\frac{1}{3}$ & $+$ \\
\hline
$\ell_L^i$ & $\mathbf{1}$ & $\mathbf{2}$ & $-\frac{1}{2}$ & $-1$ & $+$ \\
$e_R^i$ & $\mathbf{1}$ & $\mathbf{1}$ & $-1$ & $-1$ & $+$ \\
\hline
$H$ & $\mathbf{1}$ & $\mathbf{2}$ & $-\frac{1}{2}$ & $0$ & $+$ \\
\hline
$N_R^{1,2}$ & $\mathbf{1}$ & $\mathbf{1}$ & $0$ & $-1$ & $+$ \\
$\eta_L$ & $\mathbf{1}$ & $\mathbf{1}$ & $0$ & $+1$ & $-$ \\
$\xi_L$ & $\mathbf{1}$ & $\mathbf{1}$ & $0$ & $0$ & $-$ \\
\hline
$\Phi$ & $\mathbf{1}$ & $\mathbf{1}$ & $0$ & $-1$ & $+$ \\
\hline
\end{tabular}
\renewcommand{\baselinestretch}{1.1}
  \caption{Particle content of the sample $B-L$ model.} \label{BLcontent}
  \end{center}
\end{table}

The $Z^\prime$-portal framework discussed in this work is largely independent
of the underlying gauge extension of the SM.
For a concrete realization and our numerical analysis, we consider a simple
extension of the well-known minimal gauged $U(1)_{B-L}$ model
\cite{Davidson:1978pm, Mohapatra:1980qe, Marshak:1979fm, Wetterich:1981bx,
Masiero:1982fi, Mohapatra:1982xz, Buchmuller:1991ce}.
The particle content of the model is summarized in Table~\ref{BLcontent}.
In the minimal $B-L$ model, three right-handed neutrinos are introduced to
ensure anomaly cancellation, while a $B-L$ Higgs field is responsible for
the spontaneous breaking of the $U(1)_{B-L}$ gauge symmetry.
We slightly modify this setup by replacing one of the three right-handed
neutrinos with two SM-singlet fermions, $\eta_L$ and $\xi_L$, which are odd
under an imposed $Z_2$ symmetry, and introducing a $B-L$ Higgs field $\Phi$
with unit $B-L$ charge.\footnote{
A similar $U(1)_{B-L}$ symmetry-breaking structure with a Higgs field carrying
a unit $B-L$ charge was previously considered in supersymmetric
models~\cite{Okada:2016tzi, Oda:2023dqj}.}
The two right-handed neutrinos form Dirac masses with two linear combinations
of the active neutrinos, leaving one active neutrino massless. This spectrum
is sufficient to accommodate the observed neutrino oscillation data.
After $U(1)_{B-L}$ symmetry breaking, $\eta_L$ and $\xi_L$ form a pair of
nearly degenerate Majorana mass eigenstates, providing the endothermic DM
sector.

The scalar sector consists of the SM Higgs doublet $H$ and the $B-L$ Higgs
field $\Phi$. The most general renormalizable scalar potential is given by
\begin{equation}
V(H,\Phi)
=
-\mu_H^2 H^\dagger H
-\mu_\Phi^2\Phi^\dagger\Phi
+\lambda_H(H^\dagger H)^2
+\lambda_\Phi(\Phi^\dagger\Phi)^2
+\lambda_{H\Phi}(H^\dagger H)(\Phi^\dagger\Phi).
\label{scalar_potential}
\end{equation}
For simplicity, we assume that the Higgs-portal coupling
$\lambda_{H\Phi}$ is sufficiently small so that the electroweak and
$B-L$ symmetry-breaking sectors can be treated independently.

The $B-L$ Higgs field develops a vacuum expectation value (VEV),
$\langle\Phi\rangle=v_{BL}/\sqrt{2}$, spontaneously breaking
$U(1)_{B-L}$ and generating the $Z^\prime$ boson mass,
\begin{equation}
m_{Z^\prime}=g_{BL}v_{BL},
\end{equation}
where $g_{BL}$ denotes the $U(1)_{B-L}$ gauge coupling.
We parametrize the $B-L$ Higgs field around its VEV as
\begin{equation}
\Phi=\frac{1}{\sqrt{2}}\left(v_{BL}+\phi\right),
\end{equation}
where $\phi$ denotes the physical $B-L$ Higgs boson.

The electroweak symmetry is broken by
$\langle H\rangle=v_H/\sqrt{2}$, generating the SM fermion and gauge-boson
masses. In the neutrino sector, the two right-handed neutrinos
$N_R^{1,2}$ couple to the SM lepton doublets through
\begin{equation}
{\cal L}_{\nu}
\supset
-\sum_{i=1}^{3}\sum_{j=1}^{2}
Y_D^{ij}\,\overline{\ell_L^i}H N_R^j
+{\rm h.c.}
\label{Yukawa_SM}
\end{equation}
After electroweak symmetry breaking, this interaction generates a
$3\times2$ Dirac neutrino mass matrix. Consequently, two linear combinations
of the active neutrinos form massive Dirac neutrinos with $N_R^{1,2}$,
while one active neutrino remains massless. This spectrum is sufficient to
accommodate the two independent neutrino mass-squared differences observed
in neutrino oscillation experiments.

The $Z_2$-odd fermions $\eta_L$ and $\xi_L$ constitute the dark sector.
Their relevant mass and Yukawa interactions are given by
\begin{equation}
{\cal L}_{\rm dark}
\supset
-y\,\Phi\,\eta_L\xi_L
-\frac{1}{2}m_\xi\,\xi_L\xi_L
+{\rm h.c.},
\label{Yukawa_BL}
\end{equation}
where $y$ is the Yukawa coupling, and $m_\xi$ is the Majorana mass of the
$B-L$ singlet fermion $\xi_L$. 
After $U(1)_{B-L}$ symmetry breaking, the Yukawa interaction generates the Dirac mass
\begin{equation}
m_D=y \frac{v_{BL}}{\sqrt{2}},
\end{equation}
and hence, the fermion mass terms can be written as
\begin{equation}
{\cal L}_{m}
=
-\frac{1}{2}
\begin{pmatrix}
\eta_L & \xi_L
\end{pmatrix}
\begin{pmatrix}
0 & m_D\\
m_D & m_\xi
\end{pmatrix}
\begin{pmatrix}
\eta_L\\
\xi_L
\end{pmatrix}
+{\rm h.c.}
\label{dark_mass_matrix}
\end{equation}

In this work, we focus on the pseudo-Dirac limit, $m_\xi\ll m_D$.
The mass matrix in Eq.~(\ref{dark_mass_matrix}) is diagonalized by
\begin{equation}
\begin{pmatrix}
\eta_L\\
\xi_L
\end{pmatrix}
=
\begin{pmatrix}
i\cos\theta & \sin\theta\\
-i\sin\theta & \cos\theta
\end{pmatrix}
\begin{pmatrix}
\psi_{1L}\\
\psi_{2L}
\end{pmatrix},
\qquad
\tan 2\theta=\frac{2m_D}{m_\xi}.
\label{dark_mixing}
\end{equation}
The phase convention in Eq.~(\ref{dark_mixing}) is chosen such that both
mass eigenvalues are positive. In the pseudo-Dirac limit,
$\theta\simeq\pi/4$, and the masses are given by
\begin{equation}
m_{1,2}
\simeq
m_D\mp\frac{m_\xi}{2}.
\end{equation}
The two Majorana fermions are therefore nearly degenerate, with a small mass splitting \footnote{The smallness of the $m_\xi$ might be supported by the `t Hooft’s  
principle of naturalness \cite{tHooft:1979rat} since the chiral symmetry is restored in 
the limit $m_\xi \to 0$ before the $B-L$ symmetry breaking.}
\begin{equation}
\Delta m\equiv m_2-m_1\simeq m_\xi.
\label{mass_splitting}
\end{equation}
We identify the lighter state $\psi_1$ as the DM particle, while $\psi_2$
is the slightly heavier state involved in endothermic DM scattering.

For later convenience, we introduce the four-component Majorana fields
corresponding to the two mass eigenstates,
\begin{equation}
\psi_i=\psi_i^c=
\begin{pmatrix}
\psi_{iL}\\
\epsilon\,\psi_{iL}^*
\end{pmatrix},
\qquad i=1,2,
\label{Majorana_fields}
\end{equation}
where $\epsilon=i\sigma^2$. In terms of these fields, the $Z^\prime$
interactions can be conveniently expressed in the mass-eigenstate basis.

The $Z^\prime$ interaction of the dark-sector fermions originates from the
$B-L$ gauge interaction of $\eta_L$,
\begin{equation}
{\cal L}_{Z^\prime}
=
g_{BL} Z^\prime_\mu\,
\eta_L^\dagger\bar{\sigma}^\mu\eta_L .
\end{equation}
Using Eq.~(\ref{dark_mixing}), this interaction can be expressed in terms
of the four-component Majorana fields as
\begin{align}
{\cal L}_{Z^\prime}
={}&
-\frac{g_{BL}}{2}\cos^2\theta\,
Z^\prime_\mu\overline{\psi}_1\gamma^\mu\gamma^5\psi_1
-\frac{g_{BL}}{2}\sin^2\theta\,
Z^\prime_\mu\overline{\psi}_2\gamma^\mu\gamma^5\psi_2
\nonumber\\
&\quad
-i g_{BL}\sin\theta\cos\theta\,
Z^\prime_\mu\overline{\psi}_1\gamma^\mu\psi_2 .
\label{Zprime_dark_interaction}
\end{align}
In the pseudo-Dirac limit, $\theta\simeq\pi/4$, this reduces to
\begin{align}
{\cal L}_{Z^\prime}
\simeq{}&
-\frac{g_{BL}}{4}Z^\prime_\mu
\left(
\overline{\psi}_1\gamma^\mu\gamma^5\psi_1
+
\overline{\psi}_2\gamma^\mu\gamma^5\psi_2
\right)
-\frac{i g_{BL}}{2}Z^\prime_\mu
\overline{\psi}_1\gamma^\mu\psi_2 .
\label{Zprime_dark_interaction_PD}
\end{align}
Thus, the diagonal $Z^\prime$ interactions are axial-vector, whereas the
off-diagonal $\psi_1$--$\psi_2$ interaction is vector-like.
Comparing Eq.~(\ref{Zprime_dark_interaction_PD}) with the generic
$Z^\prime$-portal interaction in Eq.~(1.1), we identify
\begin{equation}
g_{12}=\frac{g_{BL}}{2},
\end{equation}
up to an irrelevant phase convention for the Majorana fields.

For completeness, the Yukawa interaction in Eq.~(\ref{Yukawa_BL}) also
induces interactions between the physical $B-L$ Higgs boson $\phi$ and
the dark-sector fermions. In the pseudo-Dirac limit, these interactions
are given by
\begin{equation}
{\cal L}_{\phi}
\simeq
-\frac{y}{2\sqrt{2}}\,\phi
\left(
\overline{\psi}_1\psi_1+
\overline{\psi}_2\psi_2
\right).
\label{phi_dark_interaction}
\end{equation}
As discussed above, we assume a sufficiently small Higgs-portal coupling
$\lambda_{H\Phi}$, so that mixing between $\phi$ and the SM Higgs boson
is negligible. Consequently, the scalar portal does not play an important
role in the DM phenomenology considered in this work. 
We therefore focus on the $Z^\prime$ portal, which directly connects the endothermic DM
scenario to $Z^\prime$ searches at the LHC.

\section{Dark Matter Phenomenology}
\label{sec:DM}

We first evaluate the DM relic abundance, taking into account the coannihilation between the nearly degenerate states $\psi_1$ and $\psi_2$. Since the mass splitting is much smaller than their masses, we neglect it in the following analysis. 
When multiple dark-sector particles are nearly degenerate in mass and
remain in chemical equilibrium during freeze-out, their coupled Boltzmann
equations can be reduced to a single equation for the total number density
$n=\sum_i n_i$. It is convenient to introduce the total yield
\begin{equation}
Y\equiv\frac{n}{s},
\end{equation}
where the entropy density is given by
\begin{equation}
s=\frac{2\pi^2}{45}g_*T^3
=\frac{2\pi^2}{45}g_*\frac{m_{\rm DM}^3}{x^3}.
\end{equation}
Here, $x\equiv m_{\rm DM}/T$ with $m_{\rm DM}=m_1$, and $g_*$ denotes
the effective number of relativistic degrees of freedom, for which we
neglect the small difference between the energy and entropy degrees of
freedom. In our numerical analysis, we use the SM value $g_*=106.75$.

The evolution of the total yield is governed by the Boltzmann equation,
\begin{equation}
\frac{dY}{dx}
=
-\frac{s\langle\sigma_{\rm eff}v_{\rm rel}\rangle}
{x^2 H(m_{\rm DM})}
\left[
Y^2-(Y^{\rm eq})^2
\right],
\label{eq:boltzmann_Y}
\end{equation}
where
\begin{equation}
H(m_{\rm DM})
=
\sqrt{\frac{4\pi^3}{45}g_*}\,
\frac{m_{\rm DM}^2}{M_{\rm Pl}},
\end{equation}
with $M_{\rm Pl}=1.22\times10^{19}~{\rm GeV}$. The equilibrium yield is
defined by
\begin{equation}
sY^{\rm eq}
=
n^{\rm eq}
=
\frac{g_{\rm eff}m_{\rm DM}^3}{2\pi^2}
\frac{K_2(x)}{x},
\label{eq:Yeq}
\end{equation}
where $K_2$ is the modified Bessel function of the second kind. Since
each Majorana fermion has two internal degrees of freedom, we have
$g_{\rm eff}=g_{\psi_1}+g_{\psi_2}=2+2=4$ in the nearly degenerate limit.

For the nearly degenerate states $\psi_1$ and $\psi_2$, the thermally
averaged effective annihilation cross section is given by
\begin{equation}
\langle\sigma_{\rm eff}v_{\rm rel}\rangle
\simeq
\frac{1}{g_{\rm eff}^{\,2}}
\sum_{i,j=1,2}g_i g_j
\langle\sigma_{ij}v_{\rm rel}\rangle.
\label{eq:sigmaeff_degenerate}
\end{equation}
Using $g_{\psi_1}=g_{\psi_2}=2$ and $g_{\rm eff}=4$, we obtain
\begin{equation}
\langle\sigma_{\rm eff}v_{\rm rel}\rangle
=
\frac{1}{16}
\left[
4\langle\sigma_{11}v_{\rm rel}\rangle
+4\langle\sigma_{22}v_{\rm rel}\rangle
+8\langle\sigma_{12}v_{\rm rel}\rangle
\right].
\label{eq:sigmaeff_pseudodirac}
\end{equation}
In our parameter region of interest, the diagonal annihilation processes
$\psi_i\psi_i\to f\bar f$, mediated by the diagonal axial-vector
$Z^\prime$ couplings, are $p$-wave suppressed in the non-relativistic
limit. They are therefore negligible compared with the off-diagonal
$\psi_1\psi_2\to f\bar f$ coannihilation process, which contains an
$s$-wave contribution. We thus approximate
\begin{equation}
\langle\sigma_{\rm eff}v_{\rm rel}\rangle
\simeq
\frac{1}{2}
\langle\sigma_{12}v_{\rm rel}\rangle.
\label{eq:sigmaeff_approx}
\end{equation}

The thermal average of the dominant coannihilation cross section is
calculated using
\begin{equation}
\left\langle \sigma_{12} v_{\rm rel} \right\rangle
=
\frac{x}{16m_{\rm DM}^5K_2^2(x)}
\int_{4m_{\rm DM}^2}^{\infty} ds\,
\hat{\sigma}_{12}(s)\sqrt{s}\,
K_1\left(\frac{x\sqrt{s}}{m_{\rm DM}}\right),
\label{avesv}
\end{equation}
where $K_1$ and $K_2$ are modified Bessel functions of the second kind,
and
\begin{equation}
\hat{\sigma}_{12}(s)
=
2\left(s-4m_{\rm DM}^2\right)\sigma_{12}(s)
\end{equation}
is the reduced coannihilation cross section.

The dominant coannihilation process is
$\psi_1\psi_2\to Z^{\prime *}\to f\bar f$, where $f$ denotes the
kinematically accessible SM fermions as well as the two right-handed
neutrinos $N_R^{1,2}$. Neglecting the SM fermion masses, the total
coannihilation cross section is given by
\begin{equation}
\sigma_{12}(s)
=
15\,
\frac{
g_{BL}^{4}\left(s+2m_{\rm DM}^{2}\right)
}{
96\pi\,\beta_{\rm DM}(s)
\left[
\left(s-m_{Z^\prime}^{2}\right)^2
+m_{Z^\prime}^{2}\Gamma_{Z^\prime}^{2}
\right]
},
\label{xsection}
\end{equation}
where
\begin{equation}
\beta_{\rm DM}(s)
=
\sqrt{1-\frac{4m_{\rm DM}^2}{s}}.
\end{equation}
Here, the factor 15 results from summing over all accessible fermion
final states, taking into account their $B-L$ charges, chiralities, and
color multiplicities.

The total decay width of the $Z^\prime$ boson, which enters the
$s$-channel propagator in Eq.~(\ref{xsection}), is given by
\begin{align}
\Gamma_{Z^\prime}
={}&
\frac{g_{BL}^2}{24\pi}m_{Z^\prime}
\bigg[
15
+\frac{1}{2}
\left(1-\frac{4m_{\rm DM}^2}{m_{Z^\prime}^2}\right)^{3/2}
\Theta\left(m_{Z^\prime}-2m_{\rm DM}\right)
\nonumber\\
&
+\frac{1}{2}
\left(1+\frac{2m_{\rm DM}^2}{m_{Z^\prime}^2}\right)
\left(1-\frac{4m_{\rm DM}^2}{m_{Z^\prime}^2}\right)^{1/2}
\Theta\left(m_{Z^\prime}-2m_{\rm DM}\right)
\bigg].
\label{eq:Zp_width}
\end{align}
Here, the first term in the square brackets represents the sum over the
SM fermions and the two right-handed neutrinos $N_R^{1,2}$. The second
and third terms arise from the diagonal $\psi_i\psi_i$ and off-diagonal
$\psi_1\psi_2$ decay channels, respectively.

Solving the Boltzmann equation, the present DM relic abundance is
determined by the asymptotic value of the yield, $Y(\infty)$, as
\begin{equation}
\Omega_{\rm DM}h^2
=
\frac{m_{\rm DM}s_0Y(\infty)}
{\rho_{\rm crit}/h^2},
\label{eq:relic_density}
\end{equation}
where $s_0=2890~{\rm cm}^{-3}$ is the present entropy density and
$\rho_{\rm crit}/h^2=1.05\times10^{-5}~{\rm GeV\,cm}^{-3}$ is the
critical density. We require the calculated relic abundance to reproduce
the observed value, $\Omega_{\rm DM}h^2\simeq0.12$~\cite{Planck:2018vyg}.

\begin{figure}[th!]
 \begin{center}
  \includegraphics[scale=0.5]{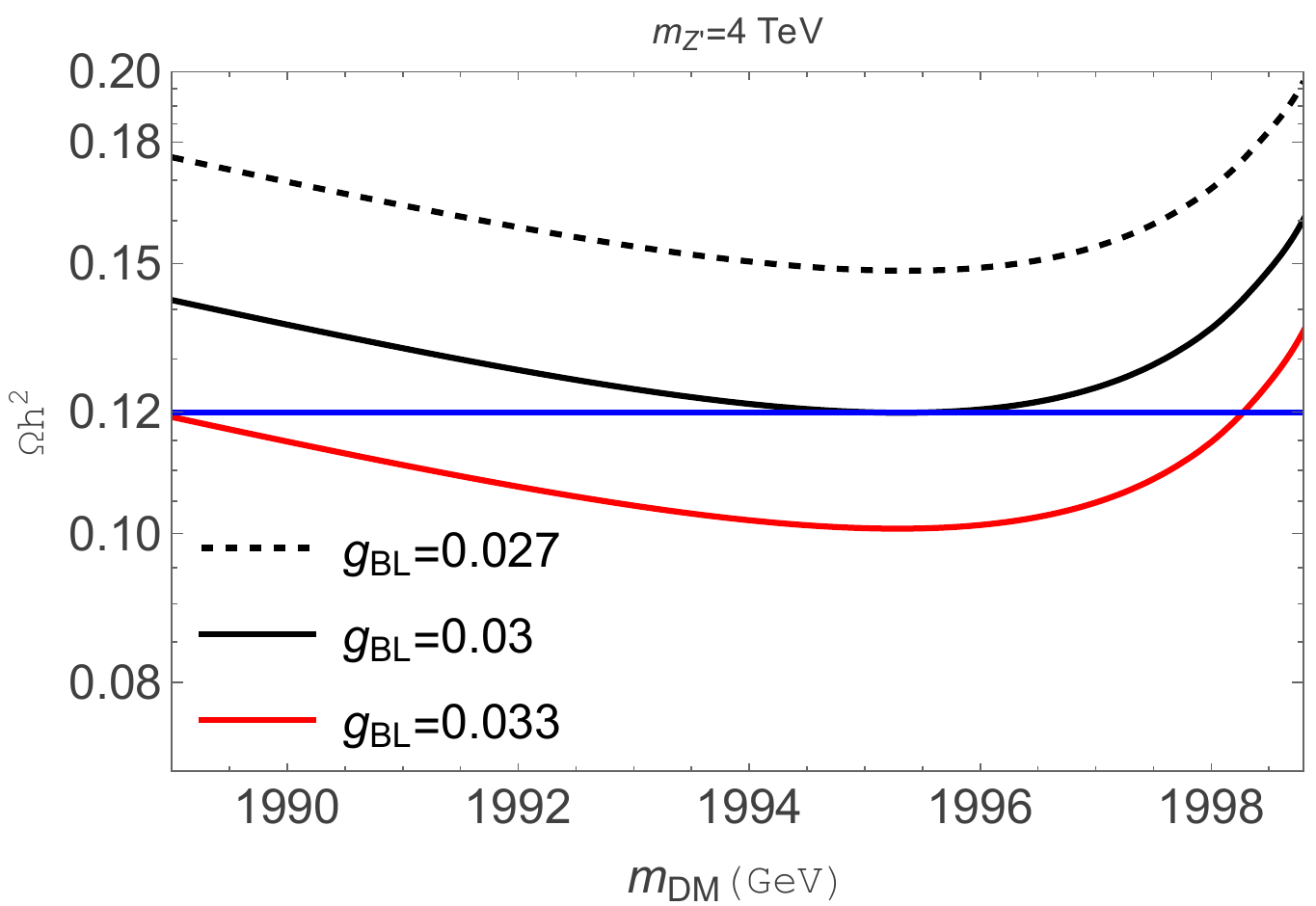}
 \end{center}
 \caption{
The relic abundance of the $Z^\prime_{B-L}$-portal endothermic Majorana
neutrino DM as a function of the DM mass, $m_{\rm DM}$, for
$m_{Z^\prime}=4~{\rm TeV}$ and $g_{BL}=0.027$, $0.030$, and $0.033$
(lines from top to bottom). The horizontal line indicates the observed
DM relic abundance, $\Omega_{\rm DM}h^2\simeq0.12$.
}
\label{BLomega}
\end{figure}

In the pseudo-Dirac limit considered here, the DM relic abundance is
essentially controlled by three free parameters, $m_{\rm DM}$,
$m_{Z^\prime}$, and $g_{BL}$, since the small mass splitting relevant
to endothermic scattering has a negligible effect on the freeze-out
calculation. Figure~\ref{BLomega} shows the DM relic abundance as a
function of $m_{\rm DM}$ for $m_{Z^\prime}=4~{\rm TeV}$ and three
representative values of the gauge coupling, $g_{BL}=0.027$, $0.030$,
and $0.033$. The relic abundance exhibits a pronounced dip near
$m_{\rm DM}\simeq m_{Z^\prime}/2$, where the coannihilation cross section
is resonantly enhanced through the $s$-channel $Z^\prime$ exchange.
For a smaller value of $g_{BL}$, even the maximal resonant enhancement
is insufficient to reduce the DM abundance to the observed value.
This allows us to determine a lower bound on $g_{BL}$ for a given
$m_{Z^\prime}$: we vary $m_{\rm DM}$ around the $Z^\prime$ resonance
and identify the minimum value of $g_{BL}$ for which
$\Omega_{\rm DM}h^2\simeq0.12$ can be reproduced. Repeating this
procedure for different values of $m_{Z^\prime}$ determines the
relic-density lower bound on $g_{BL}$ as a function of $m_{Z^\prime}$.

%

We next consider the spin-independent DM--nucleon scattering cross
section relevant to direct detection. In our model, the dominant process
is the inelastic scattering
$\psi_1 N\to\psi_2 N$, mediated by the $Z^\prime$ boson. In the limit
where the momentum transfer is much smaller than the $Z^\prime$ mass,
the spin-independent DM--nucleon scattering cross section is given by
\begin{equation}
\sigma_{\rm SI}
=
\frac{g_{BL}^{4}}{4\pi}
\frac{\mu_n^{2}}{m_{Z^\prime}^{4}},
\label{SI}
\end{equation}
where $\mu_n=\frac{m_{\rm DM}m_N}{m_{\rm DM}+m_N}$ 
is the DM--nucleon reduced mass. Since the DM mass considered in this
work is much larger than the nucleon mass, we use
$\mu_n\simeq m_N=0.938~{\rm GeV}$.

For a fixed value of $\sigma_{\rm SI}$, Eq.~(\ref{SI}) determines the
gauge coupling as a function of the $Z^\prime$ mass,
\begin{equation}
g_{BL}
=
\left(
\frac{4\pi\sigma_{\rm SI}}{\mu_n^2}
\right)^{1/4}
m_{Z^\prime}.
\label{eq:gBL_SI}
\end{equation}
Thus, a fixed value of the spin-independent DM--nucleon scattering
cross section defines a contour in the $g_{BL}$--$m_{Z^\prime}$ plane.
The value of $\sigma_{\rm SI}$ relevant to the LZ event depends
sensitively on the mass splitting $\Delta m=m_2-m_1$ through the
kinematics of endothermic scattering. For a TeV-scale DM particle,
a mass splitting of a few hundred keV is particularly relevant for
producing a nuclear recoil with $E_R\simeq250~{\rm keV}$. For
$\Delta m={\cal O}(10^2)~{\rm keV}$, the spin-independent
DM--nucleon scattering cross section relevant to the LZ event is
typically in the range
\begin{equation}
\sigma_{\rm SI}\sim10^{-46}\text{--}10^{-45}~{\rm cm}^2,
\label{eq:SI_range}
\end{equation}
depending on the mass splitting and the DM mass; see, for example,
Ref.~\cite{Borah:2026zwf}. We therefore consider such representative values
of $\sigma_{\rm SI}$ in this range and translate them into contours in
the $g_{BL}$--$m_{Z^\prime}$ plane using Eq.~(\ref{eq:gBL_SI}). 
These contours will be shown in the next section, together with the
relic-density requirement and the constraint from the LHC $Z^\prime$ resonance search.

%
\begin{figure}[tb]
 \begin{center}
  \includegraphics[scale=0.55]{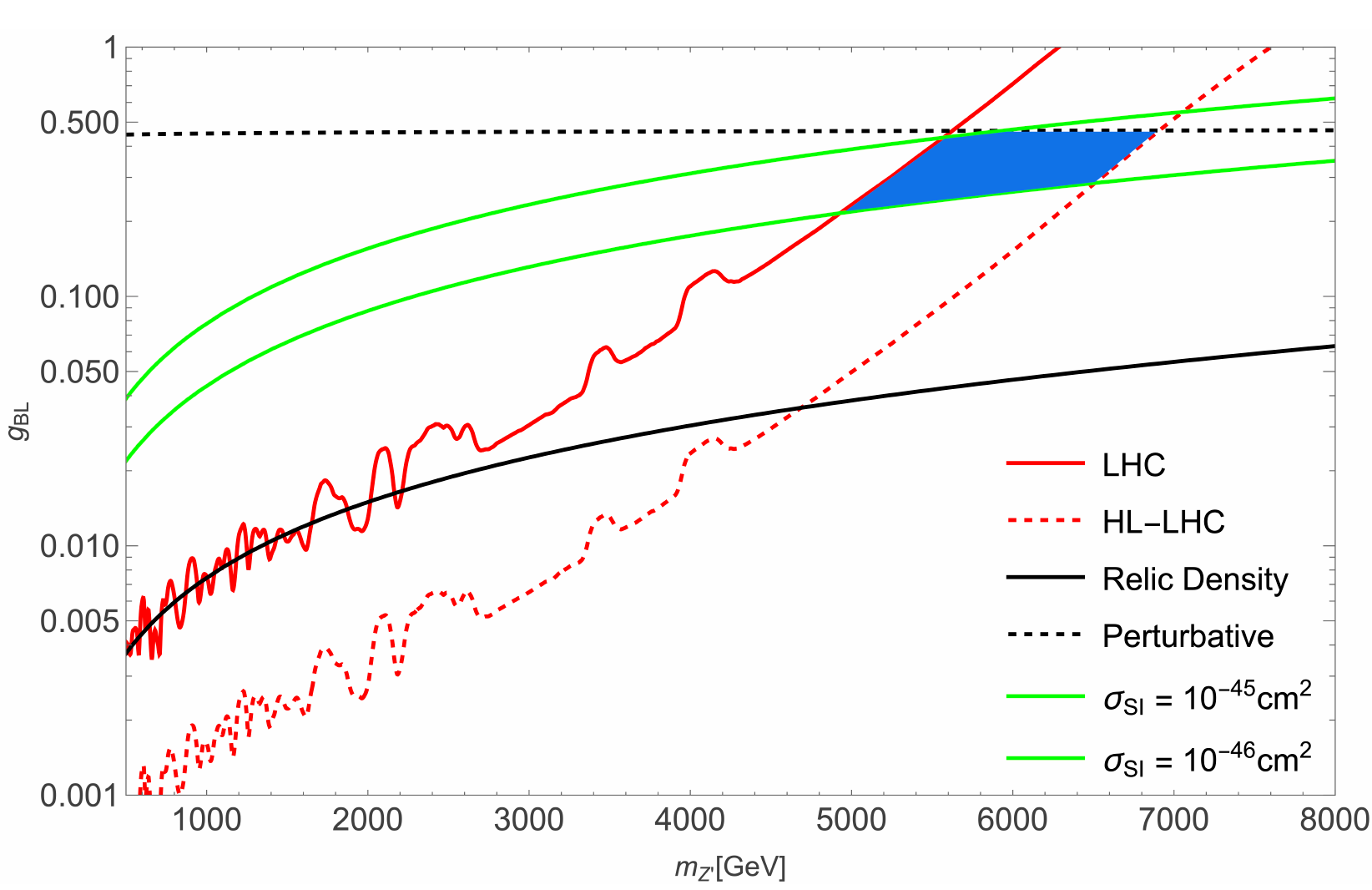}
 \end{center}
 \renewcommand{\baselinestretch}{1.1}
 \caption{ 
Allowed parameter region for the $Z^\prime_{B-L}$-portal endothermic
Majorana DM scenario in the $g_{BL}$--$m_{Z^\prime}$ plane.
The solid black curve represents the lower bound on $g_{BL}$ required
to reproduce the observed DM relic abundance. The solid red curve
shows the upper bound from the ATLAS $Z^\prime$ dilepton resonance
search~\cite{ATLAS:2019erb}, while the dashed red curve represents the
estimated sensitivity of the HL-LHC. The solid green lines corresponds to spin-independent DM--nucleon scattering cross sections $\sigma_{\rm SI}=10^{-45}~{\rm cm}^2$ (top) and $10^{-46}~{\rm cm}^2$ (bottom) relevant to the LZ event.
The solid black dashed curve denotes the theoretical upper bound on $g_{BL}$
obtained by requiring the absence of a Landau pole below $M_{\rm Pl}$. The blue shaded region is bounded by requiring the parameters to reproduce the observed DM relic abundance and the LZ event while satisfying perturbativity and current ATLAS $Z^\prime$ dilepton resonance search bounds, 
alongside the projected HL-LHC search reach for the $Z^\prime$ resonance. }
\label{BLfinal}
\end{figure}
%

\section{LHC Constraints and Complementarity}
\label{sec:LHC}

We now consider constraints on the $Z^\prime$ boson from resonance
searches at the LHC and examine their complementarity with the DM
constraints discussed in the previous section. In particular, searches
for a high-mass dilepton resonance~\cite{ATLAS:2019erb,CMS:2021ctt}
provide a stringent upper bound on the $B-L$ gauge coupling $g_{BL}$
as a function of $m_{Z^\prime}$. We evaluate the resonant dilepton
($\ell=e,\mu$) production cross section at $\sqrt{s}=13~{\rm TeV}$.
Since the values of $g_{BL}$ relevant to our analysis are relatively
small, the $Z^\prime$ boson has a narrow decay width, and we employ
the narrow-width approximation (NWA) in evaluating the production
cross section.

The inclusive $Z^\prime$ production cross section at the LHC is
calculated from the quark--antiquark annihilation process
$q\bar q\to Z^\prime$ as
\begin{equation}
\sigma(pp\to Z^\prime)
=
\sum_q
\int_0^1 dx_1\int_0^1 dx_2\,
\left[
f_q(x_1,Q)f_{\bar q}(x_2,Q)
+
f_{\bar q}(x_1,Q)f_q(x_2,Q)
\right]
\hat{\sigma}_{q\bar q}(\hat{s}),
\label{eq:Zprime_production}
\end{equation}
where $\hat{s}=x_1x_2s$ is the partonic center-of-mass energy squared,
and $f_q(x,Q)$ and $f_{\bar q}(x,Q)$ denote the quark and antiquark
parton distribution functions (PDFs), respectively. In the NWA, the
partonic cross section is given by
\begin{equation}
\hat{\sigma}_{q\bar q}(\hat{s})
=
\frac{4\pi^2}{3}
\frac{\Gamma(Z^\prime\to q\bar q)}{m_{Z^\prime}}
\delta\left(\hat{s}-m_{Z^\prime}^2\right).
\label{eq:Zprime_NWA}
\end{equation}
The dilepton signal cross section is then obtained as
\begin{equation}
\sigma(pp\to Z^\prime\to\ell^+\ell^-)
=
\sigma(pp\to Z^\prime)\,
{\rm BR}(Z^\prime\to\ell^+\ell^-),
\qquad \ell=e,\mu .
\label{eq:Zprime_dilepton}
\end{equation}

For the PDFs, we employ the CTEQ6L set~\cite{Pumplin:2002vw}, with the
factorization scale fixed at $Q=m_{Z^\prime}$. We compare the predicted
dilepton production cross section,
$\sigma(pp\to Z^\prime\to\ell^+\ell^-)$, with the upper limits from the
ATLAS Run-2 dilepton resonance search~\cite{ATLAS:2019erb}. The ATLAS
limits are then translated into an upper bound on $g_{BL}$ as a function
of $m_{Z^\prime}$. The resulting LHC Run-2 constraint is shown by the
red solid curve in Fig.~\ref{BLfinal}.

We also estimate the sensitivity of the HL-LHC to the $Z^\prime$
dilepton resonance. In the high-mass region relevant to our analysis,
the SM background is expected to be very small, and we therefore
estimate the HL-LHC sensitivity by scaling the Run-2 result with the
integrated luminosity. Since the $Z^\prime$ production cross section
is proportional to $g_{BL}^2$, the projected upper bound on the gauge
coupling scales as $\mathcal{L}^{-1/2}$. Taking the Run-2 integrated
luminosity to be $137~{\rm fb}^{-1}$ and the HL-LHC target luminosity
to be $3000~{\rm fb}^{-1}$, we obtain
\begin{equation}
g_{BL}^{\rm HL-LHC}
\simeq
g_{BL}^{\rm Run\text{-}2}
\sqrt{\frac{137}{3000}}.
\label{eq:HLLHC_scaling}
\end{equation}
The resulting HL-LHC sensitivity is shown by the red dashed curve in
Fig.~\ref{BLfinal}.

We also impose the theoretical requirement that the $U(1)_{B-L}$
gauge coupling remain perturbative up to the Planck scale. At the
one-loop level, the running of the gauge coupling is governed by
\begin{equation}
\frac{d g_{BL}}{d\ln\mu}
=
\frac{11}{16\pi^2}g_{BL}^3.
\end{equation}
Requiring the absence of a Landau pole below $M_{\rm Pl}$ leads to
the upper bound
\begin{equation}
g_{BL}(m_{Z^\prime})
<
\sqrt{
\frac{8\pi^2}
{11\ln\left(M_{\rm Pl}/m_{Z^\prime}\right)}
}.
\label{eq:Landau}
\end{equation}
Here, $g_{BL}$ used in our DM and collider analyses is understood as
the running gauge coupling evaluated at $\mu=m_{Z^\prime}$. The
resulting upper bound is shown by the black dashed curve in
Fig.~\ref{BLfinal}.

We also show in Fig.~\ref{BLfinal} contours of fixed spin-independent
DM--nucleon scattering cross sections relevant to the LZ event.
As discussed in the previous section, for a fixed value of
$\sigma_{\rm SI}$, Eq.~(\ref{eq:gBL_SI}) determines $g_{BL}$ as a
function of $m_{Z^\prime}$. As representative values, we take
$\sigma_{\rm SI}=10^{-45}~{\rm cm}^2$ and
$10^{-46}~{\rm cm}^2$, which is relevant to the endothermic DM interpretation of the LZ event 
with a mass splitting of around $100$ keV. 
The corresponding contours are shown by green lines from top to bottom, respectively, in Fig.~\ref{BLfinal}. 
The region between the two lines illustrates the parameter region in which the interaction strength relevant to the LZ event can be realized in our $Z^\prime$-portal model.

Figure~\ref{BLfinal} summarizes the complementarity between the DM
relic-density requirement and the LHC $Z^\prime$ resonance searches.
For a given $m_{Z^\prime}$, the relic-density requirement imposes a
lower bound on $g_{BL}$, since a sufficiently large gauge coupling is
required to achieve the observed DM abundance even with the resonant
enhancement of the coannihilation cross section. On the other hand,
the LHC dilepton resonance search imposes an upper bound on $g_{BL}$.
The parameter region between these two bounds therefore simultaneously
satisfies the DM relic-density requirement and the current LHC
constraint. Remarkably, the representative $\sigma_{\rm SI}$ contours
relevant to the LZ event pass through this allowed region, demonstrating
the complementarity among the LZ event, thermal DM, and collider
searches. The projected HL-LHC sensitivity will further probe this
parameter region. 
The blue shaded region denotes the parameter space that reproduces the observed DM relic abundance and the LZ event while satisfying perturbativity and current ATLAS $Z^\prime$ dilepton resonance search bounds. 
The region is bounded on the right by the projected HL-LHC search reach for the $Z^\prime$ resonance.

\section{Conclusion}

We have investigated an endothermic DM interpretation of the high-energy
nuclear-recoil event reported by the LZ Collaboration in a simple
$Z^\prime$-portal framework. As a concrete realization, we considered
a gauged $U(1)_{B-L}$ model in which the DM sector consists of two
nearly degenerate Majorana fermions. A mass splitting of
${\cal O}(10^2)~{\rm keV}$ allows the lighter state to undergo
endothermic inelastic scattering off nuclei, while the observed DM
relic abundance can be reproduced through resonantly enhanced
coannihilation near $m_{\rm DM}\simeq m_{Z^\prime}/2$.

An important feature of this scenario is the complementarity between
DM and collider phenomenology. The relic-density requirement provides
a lower bound on the $B-L$ gauge coupling $g_{BL}$, whereas LHC
dilepton resonance searches impose an upper bound. We have shown that
spin-independent scattering cross sections relevant to the LZ event
can be realized in the parameter region between these bounds.
The HL-LHC will further probe this region, providing a direct test of
the $Z^\prime$-portal interpretation of the LZ event.

\section*{Acknowledgments}
The work of N.O. is supported in part by the United States Department of Energy Grants No.~DC-SC0023713.

\newpage
\bibliographystyle{utphysII}
\bibliography{References}

@inproceedings{tHooft:1979rat,
    author = "'t Hooft, Gerard",
    title = "{Naturalness, chiral symmetry, and spontaneous chiral symmetry breaking}",
    booktitle = "{Cargese Summer Institute: Recent Developments in Gauge Theories}",
    volume = "59",
    pages = "135--157",
    year = "1980",
    doi = "10.1007/978-1-4684-7571-5_9"
}

@article{Das:2026buc,
    author = "Das, Arindam and Nomura, Takaaki",
    title = "{Effect of inelastic scalar dark matter in hidden $U(1)$ scenario after the LZ nuclear recoil}",
    eprint = "2609.15600",
    archivePrefix = "arXiv",
    primaryClass = "hep-ph",
    month = "9",
    year = "2026"
}

@article{Okada:2019sbb,
    author = "Okada, Nobuchika and Seto, Osamu",
    title = "{Inelastic extra $U(1)$ charged scalar dark matter}",
    eprint = "1908.09277",
    archivePrefix = "arXiv",
    primaryClass = "hep-ph",
    reportNumber = "EPHOU-19-011",
    doi = "10.1103/PhysRevD.101.023522",
    journal = "Phys. Rev. D",
    volume = "101",
    number = "2",
    pages = "023522",
    year = "2020"
}

@article{Mahapatra:2026glu,
    author = "Mahapatra, Satyabrata and Paul, Partha Kumar",
    title = "{Boosted or Inelastic? Discriminating Interpretations of the LZ 248 keV Event}",
    eprint = "2609.14799",
    archivePrefix = "arXiv",
    primaryClass = "hep-ph",
    month = "9",
    year = "2026"
}

@article{Palmisano:2026kuj,
    author = "Palmisano, Stefano and Tammaro, Michele and Tesi, Andrea",
    title = "{Inferring dark matter masses and interactions from high recoil energy events in LUX-ZEPLIN}",
    eprint = "2609.15985",
    archivePrefix = "arXiv",
    primaryClass = "hep-ph",
    month = "9",
    year = "2026"
}

@article{Bose:2026szs,
    author = "Bose, Debajit and Saha, Akash Kumar and Raj, Nirmal and Maity, Tarak Nath and Laha, Ranjan",
    title = "{LUX-ZEPLIN's Stairway to Hea$\nu$en: limits on elastic scatters of dark matter from solar capture}",
    eprint = "2609.21823",
    archivePrefix = "arXiv",
    primaryClass = "hep-ph",
    month = "9",
    year = "2026"
}

@article{He:2026idw,
    author = "He, Xiao-Gang and Hong, Xuan and Jeesun, Sk",
    title = "{Hadrophilic inelastic freeze-in dark matter in $q_1-q_2$ gauge extension and the high energy LZ event}",
    eprint = "2609.15714",
    archivePrefix = "arXiv",
    primaryClass = "hep-ph",
    month = "9",
    year = "2026"
}

@article{Unwin:2026rdp,
    author = "Unwin, James",
    title = "{Axion Portal Dark Matter and the LUX-ZEPLIN High-Recoil Event}",
    eprint = "2609.04186",
    archivePrefix = "arXiv",
    primaryClass = "hep-ph",
    month = "9",
    year = "2026"
}

@article{Borah:2026ris,
    author = "Borah, Pankaj and Mahapatra, Satyabrata and Nath, Newton",
    title = "{Inelastic Dark Matter at LZ from Radiative Dirac Neutrino Mass Paradigm}",
    eprint = "2609.15027",
    archivePrefix = "arXiv",
    primaryClass = "hep-ph",
    month = "9",
    year = "2026"
}

@article{Davidson:1978pm,

    author = "Davidson, Aharon",

    title = "{$B-L$ as the fourth color within an $\mathrm{SU}(2)_L \times  \mathrm{U}(1)_R \times \mathrm{U}(1)$ model}",

    reportNumber = "SU-4213-129, COO-3533-129",

    doi = "10.1103/PhysRevD.20.776",

    journal = "Phys. Rev. D",

    volume = "20",

    pages = "776",

    year = "1979"

}

@article{Mohapatra:1980qe,

    author = "Mohapatra, Rabindra N. and Marshak, R. E.",

    title = "{Local B-L Symmetry of Electroweak Interactions, Majorana Neutrinos and Neutron Oscillations}",

    reportNumber = "VPI-HEP-80/1",

    doi = "10.1103/PhysRevLett.44.1316",

    journal = "Phys. Rev. Lett.",

    volume = "44",

    pages = "1316--1319",

    year = "1980",

    note = "[Erratum: Phys.Rev.Lett. 44, 1643 (1980)]"

}

@article{Marshak:1979fm,

    author = "Marshak, R. E. and Mohapatra, Rabindra N.",

    title = "{Quark - Lepton Symmetry and B-L as the U(1) Generator of the Electroweak Symmetry Group}",

    reportNumber = "Print-80-0096 (VIRGINIA TECH)",

    doi = "10.1016/0370-2693(80)90436-0",

    journal = "Phys. Lett. B",

    volume = "91",

    pages = "222--224",

    year = "1980"

}

@article{Wetterich:1981bx,

    author = "Wetterich, C.",

    title = "{Neutrino Masses and the Scale of B-L Violation}",

    reportNumber = "FREIBURG-THEP-81-2",

    doi = "10.1016/0550-3213(81)90279-0",

    journal = "Nucl. Phys. B",

    volume = "187",

    pages = "343--375",

    year = "1981"

}

@article{Masiero:1982fi,

    author = "Masiero, A. and Nieves, J. F. and Yanagida, T.",

    title = "{$B-L$ Violating Proton Decay and Late Cosmological Baryon Production}",

    reportNumber = "MPI-PAE/PTh 26/82",

    doi = "10.1016/0370-2693(82)90024-7",

    journal = "Phys. Lett. B",

    volume = "116",

    pages = "11--15",

    year = "1982"

}

@article{Mohapatra:1982xz,

    author = "Mohapatra, Rabindra N. and Senjanovic, Goran",

    title = "{Spontaneous Breaking of Global $B-L$ Symmetry and Matter - Antimatter Oscillations in Grand Unified Theories}",

    reportNumber = "BNL-31720, IC/82/89",

    doi = "10.1103/PhysRevD.27.254",

    journal = "Phys. Rev. D",

    volume = "27",

    pages = "254",

    year = "1983"

}

@article{Buchmuller:1991ce,

    author = "Buchmuller, W. and Greub, C. and Minkowski, P.",

    title = "{Neutrino masses, neutral vector bosons and the scale of B-L breaking}",

    reportNumber = "DESY-91-053",

    doi = "10.1016/0370-2693(91)90952-M",

    journal = "Phys. Lett. B",

    volume = "267",

    pages = "395--399",

    year = "1991"

}

@article{LZ:2026axp,
    author = "Akerib, D. S. and others",
    collaboration = "LZ",
    title = "{Search for dark matter particle interactions in an extended nuclear recoil energy window with the LUX-ZEPLIN (LZ) experiment}",
    eprint = "2609.02823",
    archivePrefix = "arXiv",
    primaryClass = "hep-ex",
    doi = "10.17182/hepdata.182472.v1",
    month = "9",
    year = "2026"
}

@article{Baer:2026fpy,
    author = "Baer, Howard and Barger, Vernon",
    title = "{Exothermic dark matter and the 248 keV nuclear recoil in LUX-ZEPLIN}",
    eprint = "2609.06153",
    archivePrefix = "arXiv",
    primaryClass = "hep-ph",
    month = "9",
    year = "2026"
}

@article{deLima:2026shq,
    author = "de Lima, Carlos Henrique",
    title = "{Exothermic Dark Matter at LZ}",
    eprint = "2609.05204",
    archivePrefix = "arXiv",
    primaryClass = "hep-ph",
    month = "9",
    year = "2026"
}

@article{Dent:2026bji,
    author = "Dent, James B. and Newstead, Jayden L.",
    title = "{Exothermic and Endothermic Inelastic Dark Matter Interpretations at LZ: Sideband Constraints and Future Prospects}",
    eprint = "2609.04673",
    archivePrefix = "arXiv",
    primaryClass = "hep-ph",
    month = "9",
    year = "2026"
}

@article{Fan:2026hzw,
    author = "Fan, Zi-Tong and He, Hong-Jian and Wang, Yu-Chen and Zhao, Yue",
    title = "{Inelastic Dark Matter and High-Energy Recoil Signatures in LZ}",
    eprint = "2609.10491",
    archivePrefix = "arXiv",
    primaryClass = "hep-ph",
    month = "9",
    year = "2026"
}

@article{Alhazmi:2026efz,
    author = "Alhazmi, Haider and Kim, Doojin and Kong, Kyoungchul and Park, Jong-Chul and Shin, Seodong",
    title = "{High-Energy Nuclear Recoils from Boosted Dark Matter for the LZ 248-keV Event: Beyond the Halo-Dependent High-Velocity Tail}",
    eprint = "2609.06890",
    archivePrefix = "arXiv",
    primaryClass = "hep-ph",
    month = "9",
    year = "2026"
}

@article{Kannike:2026qyl,
    author = "Kannike, Kristjan and Raidal, Martti and Strumia, Alessandro",
    title = "{Boosted dark particles and the LZ nuclear recoil event}",
    eprint = "2609.07742",
    archivePrefix = "arXiv",
    primaryClass = "hep-ph",
    month = "9",
    year = "2026"
}

@article{Liang:2026coz,
    author = "Liang, Jin-Han and Liu, Zuowei and Tran, Van Que and Xu, Yongheng",
    title = "{LZ Nuclear-Recoil Excess from Boosted Light Magnetic Dipole-dipole Dark Matter}",
    eprint = "2609.06756",
    archivePrefix = "arXiv",
    primaryClass = "hep-ph",
    month = "9",
    year = "2026"
}

@article{Heikinheimo:2026kwp,
    author = "Heikinheimo, Matti and Zimmermann, Niklas",
    title = "{Cosmic ray boosted dark matter with momentum dependent interactions can explain the LZ 248 keV event}",
    eprint = "2609.11600",
    archivePrefix = "arXiv",
    primaryClass = "hep-ph",
    month = "9",
    year = "2026"
}

@article{Lou:2026idn,
    author = "Lou, Yuanchao and Lu, Chih-Ting",
    title = "{Fermionic Dark Matter Absorption and the High-Energy Event in LUX-ZEPLIN}",
    eprint = "2609.01592",
    archivePrefix = "arXiv",
    primaryClass = "hep-ph",
    month = "9",
    year = "2026"
}

@article{Jeesun:2026vzo,
    author = "Jeesun, Sk and Majumdar, Anirban",
    title = "{Atmospheric neutrino up-scattering explanation of LZ 2026 excess}",
    eprint = "2609.04185",
    archivePrefix = "arXiv",
    primaryClass = "hep-ph",
    month = "9",
    year = "2026"
}

@article{Chattaraj:2026fxn,
    author = "Chattaraj, Ayan and Majumdar, Anirban and Papoulias, Dimitrios K. and Srivastava, Rahul",
    title = "{Can Elastic Neutrino Scattering Account for the LZ230616 Event?}",
    eprint = "2609.10504",
    archivePrefix = "arXiv",
    primaryClass = "hep-ph",
    month = "9",
    year = "2026"
}

@article{Aghaie:2026vsu,
    author = "Aghaie, Mohammad and Strumia, Alessandro",
    title = "{Neutron disappearance and the LZ nuclear recoil event}",
    eprint = "2609.09037",
    archivePrefix = "arXiv",
    primaryClass = "hep-ph",
    month = "9",
    year = "2026"
}

@article{Lee:2026zbr,
    author = "Lee, Junseok and Takahashi, Fuminobu and Tsai, Yu-Dai",
    title = "{Nuclear Recoils from Invisible Neutron-Pair Annihilation and the LZ event}",
    eprint = "2609.12045",
    archivePrefix = "arXiv",
    primaryClass = "hep-ph",
    month = "9",
    year = "2026"
}

@article{Gu:2026vto,
    author = "Gu, Guanhua and Li, Lingfeng and Tang, Shao-Song and Xu, Yongheng",
    title = "{Inelastic from the Other Side: Xenon Excitation Signals in Light of the LZ High-Recoil Event}",
    eprint = "2609.05291",
    archivePrefix = "arXiv",
    primaryClass = "hep-ph",
    month = "9",
    year = "2026"
}

@article{An:2026pkc,
    author = "An, Haipeng and Gao, Fei and Liu, Jia and Liu, Minghao and Xu, Changlong",
    title = "{Cosmological Constrained Axion-Portal Inelastic Dark Matter for the LZ Event}",
    eprint = "2609.17412",
    archivePrefix = "arXiv",
    primaryClass = "hep-ph",
    month = "9",
    year = "2026"
}

@article{Fan:2026kxx,
    author = "Fan, JiJi and Reece, Matthew",
    title = "{Higgsino Above the Sea of Fog}",
    eprint = "2609.01504",
    archivePrefix = "arXiv",
    primaryClass = "hep-ph",
    month = "9",
    year = "2026"
}

@article{Freese:2026sga,
    author = "Freese, Katherine and Theodosopoulos, Dionysios P.",
    title = "{Higgsino Dark Matter Interpretation of the LUX-ZEPLIN 248 keV Nuclear-Recoil Event}",
    eprint = "2609.01583",
    archivePrefix = "arXiv",
    primaryClass = "hep-ph",
    month = "9",
    year = "2026"
}

@article{Wu:2026nhi,
    author = "Wu, Lei and Zhang, Yang and Zhu, Bin",
    title = "{TeV Higgsino Dark Matter from LZ Nuclear Recoil to Fermi-LAT Gamma Rays}",
    eprint = "2609.01590",
    archivePrefix = "arXiv",
    primaryClass = "hep-ph",
    month = "9",
    year = "2026"
}

@article{Yin:2026jnn,
    author = "Yin, Wen",
    title = "{A PQ-Symmetric High-Scale SUSY Interpretation of the LZ High-Energy Recoil}",
    eprint = "2609.01892",
    archivePrefix = "arXiv",
    primaryClass = "hep-ph",
    month = "9",
    year = "2026"
}

@article{Du:2026guj,
    author = "Du, Xiaokang and Wang, Fei",
    title = "{TeV Higgsino Interpretation of the LZ High-Recoil Event with Intermediate-Scale Electroweak Gauginos}",
    eprint = "2609.04163",
    archivePrefix = "arXiv",
    primaryClass = "hep-ph",
    month = "9",
    year = "2026"
}

@article{Langhoff:2026ujr,
    author = "Langhoff, Kevin",
    title = "{Heavy Higgsino Interpretation of the LZ Event}",
    eprint = "2609.09385",
    archivePrefix = "arXiv",
    primaryClass = "hep-ph",
    month = "9",
    year = "2026"
}

@article{Bisal:2026khf,
    author = "Bisal, Subhadip and Cao, Junjie and Li, Fei",
    title = "{Higgsino Dark Matter Interpretation of the LZ High-Recoil Event in the GNMSSM with TeV-Scale Gauginos}",
    eprint = "2609.07811",
    archivePrefix = "arXiv",
    primaryClass = "hep-ph",
    month = "9",
    year = "2026"
}

@article{Visinelli:2026kgt,
    author = "Visinelli, Luca",
    title = "{A Peccei-Quinn Origin for Inelastic Electroweak Dark Matter after LUX-ZEPLIN}",
    eprint = "2609.02807",
    archivePrefix = "arXiv",
    primaryClass = "hep-ph",
    month = "9",
    year = "2026"
}

@article{Smirnov:2026aqk,
    author = "Smirnov, Juri and Griffith, Spencer and Beacom, John F.",
    title = "{Inelastic Signatures of Electroweak Dark Matter}",
    eprint = "2609.04144",
    archivePrefix = "arXiv",
    primaryClass = "hep-ph",
    month = "9",
    year = "2026"
}

@article{Nomura:2026qyq,
    author = "Nomura, Yasunori",
    title = "{Dark Matter as the Z{\_}2 Partner of the Standard Model Higgs Boson}",
    eprint = "2609.02505",
    archivePrefix = "arXiv",
    primaryClass = "hep-ph",
    reportNumber = "RIKEN-iTHEMS-Report-26",
    month = "9",
    year = "2026"
}

@article{Wang:2026ytg,
    author = "Wang, Lei and Xiao, Yang",
    title = "{The Inert Doublet Model of Dark Matter and the LUX-ZEPLIN High-Recoil Event}",
    eprint = "2609.06571",
    archivePrefix = "arXiv",
    primaryClass = "hep-ph",
    month = "9",
    year = "2026"
}

@article{Su:2026rwz,
    author = "Su, Liangliang and Yang, Jin Min and Yang, Wen-Na",
    title = "{Inelastic Dark Matter Signature at High Recoil Energy in LUX-ZEPLIN and CRESST}",
    eprint = "2609.01475",
    archivePrefix = "arXiv",
    primaryClass = "hep-ph",
    month = "9",
    year = "2026"
}

@article{McCabe:2026crm,
    author = "McCabe, Christopher",
    title = "{Seasonal dark matter from the LUX-ZEPLIN high-energy event}",
    eprint = "2609.04181",
    archivePrefix = "arXiv",
    primaryClass = "hep-ph",
    month = "9",
    year = "2026"
}

@article{DiMauro:2026ldr,
    author = "Di Mauro, Mattia",
    title = "{Dark Matter at the Kinematic Edge: Interpreting the 248 keV LZ Nuclear-Recoil Candidate}",
    eprint = "2609.02608",
    archivePrefix = "arXiv",
    primaryClass = "hep-ph",
    month = "9",
    year = "2026"
}

@article{Yamashita:2026ump,
    author = "Yamashita, Kimiko",
    title = "{Inelastic Dark Photon Dark Matter for the LUX-ZEPLIN High-Recoil Event and the Galactic Halo Gamma-Ray Excess}",
    eprint = "2609.02868",
    archivePrefix = "arXiv",
    primaryClass = "hep-ph",
    month = "9",
    year = "2026"
}

@article{Lee:2026wof,
    author = "Lee, Hyun Min",
    title = "{Inelastic dark matter and baryon flavor symmetry in light of LUX-ZEPLIN (LZ) experiment}",
    eprint = "2609.06171",
    archivePrefix = "arXiv",
    primaryClass = "hep-ph",
    month = "9",
    year = "2026"
}

@article{Das:2026uyy,
    author = "Das, Pritam and Karmakar, Biswajit and Mahapatra, Satyabrata and Paul, Partha Kumar",
    title = "{Inelastic Self-interacting Dark Matter and LUX-ZEPLIN 248 keV Event in a Dirac Modular Inverse Seesaw}",
    eprint = "2609.06825",
    archivePrefix = "arXiv",
    primaryClass = "hep-ph",
    month = "9",
    year = "2026"
}

@article{Okada:2026eol,
    author = "Okada, Nobuchika and Seto, Osamu",
    title = "{Inelastic $B-L$ scalar dark matter and the LUX-ZEPLIN event}",
    eprint = "2609.06909",
    archivePrefix = "arXiv",
    primaryClass = "hep-ph",
    reportNumber = "EPHOU-26-011",
    month = "9",
    year = "2026"
}

@article{Kumar:2026lgi,
    author = "Kumar, Ranjeet and Prajapati, Hemant Kumar",
    title = "{Generalized Chiral $U(1)_{B-L}$ with Inelastic Scalar Dark Matter for the LZ 248 keV Event}",
    eprint = "2609.10827",
    archivePrefix = "arXiv",
    primaryClass = "hep-ph",
    month = "9",
    year = "2026"
}

@article{Bandyopadhyay:2026gjw,
    author = "Bandyopadhyay, Disha and Borah, Debasish and Borah, Pankaj",
    title = "{LZ nuclear recoil event from inelastic singlet-doublet scalar dark matter}",
    eprint = "2609.07451",
    archivePrefix = "arXiv",
    primaryClass = "hep-ph",
    month = "9",
    year = "2026"
}

@article{Borah:2026zwf,
    author = "Borah, Debasish and Sahoo, Sujit Kumar and Sahu, Narendra and Sharma, Shashwat",
    title = "{Inelastic Singlet-Doublet Fermion Dark Matter in light of the 248 keV LZ event}",
    eprint = "2609.07800",
    archivePrefix = "arXiv",
    primaryClass = "hep-ph",
    month = "9",
    year = "2026"
}

@article{Du:2026lpa,
    author = "Du, Xin-Yu and Huang, Wenjie and Xie, Keping",
    title = "{Pseudo-Dirac Inelastic Dark Matter in the Leptophobic $U(1)_B$ Model: Confronting the LUX-ZEPLIN High-Recoil Event with Collider Searches}",
    eprint = "2609.07225",
    archivePrefix = "arXiv",
    primaryClass = "hep-ph",
    month = "9",
    year = "2026"
}

@article{Yang:2026wpb,
    author = "Yang, Meiwen and Wu, Quan-feng and Tsai, Yue-Lin Sming and Fan, Yi-Zhong",
    title = "{Multi-Messenger and Paleo-Detector Probes of the LZ Dark Matter Signal}",
    eprint = "2609.06640",
    archivePrefix = "arXiv",
    primaryClass = "hep-ph",
    month = "9",
    year = "2026"
}

@article{Ahmed:2026qjg,
    author = "Ahmed, Waqas and Leontaris, George K.",
    title = "{A Dark-Dimension Origin of Geometric Inelastic Dark Matter: The LUX-ZEPLIN High-Recoil Event and Multi-Target Tests}",
    eprint = "2609.07138",
    archivePrefix = "arXiv",
    primaryClass = "hep-ph",
    month = "9",
    year = "2026"
}

@article{Lee:2026xxh,
    author = "Lee, Vincent S. H. and Randall, Lisa",
    title = "{A Warped Extra Dimensional Candidate for the LZ 248 keV Event}",
    eprint = "2609.09136",
    archivePrefix = "arXiv",
    primaryClass = "hep-ph",
    reportNumber = "N3AS-26-020",
    month = "9",
    year = "2026"
}

@article{Lee:2026jxl,
    author = "Lee, Seung J. and Youn, Taewook",
    title = "{Mixing-suppressed inelastic dark matter: a minimal model for the LZ 248 keV event}",
    eprint = "2609.09138",
    archivePrefix = "arXiv",
    primaryClass = "hep-ph",
    month = "9",
    year = "2026"
}

@article{Zhu:2026dag,
    author = "Zhu, Pengxuan and Dalla Valle Garcia, Giovani and Wang, Xuan-Gong and Thomas, Anthony W. and White, Martin J.",
    title = "{Endothermic dark matter with a light dark photon and the LUX--ZEPLIN high-energy nuclear-recoil candidate}",
    eprint = "2609.09015",
    archivePrefix = "arXiv",
    primaryClass = "hep-ph",
    month = "9",
    year = "2026"
}

@article{Qi:2026vyp,
    author = "Qi, XinXin and Sun, Hao",
    title = "{Nonthermal Solar Stalling of an Inelastic Scalar Signal in Xenon}",
    eprint = "2609.10636",
    archivePrefix = "arXiv",
    primaryClass = "hep-ph",
    month = "9",
    year = "2026"
}

@article{He:2026hqz,
    author = "He, Yuxuan",
    title = "{Transition magnetic-dipole dark matter and the LZ230616 high-recoil candidate}",
    eprint = "2609.10453",
    archivePrefix = "arXiv",
    primaryClass = "hep-ph",
    month = "9",
    year = "2026"
}

@article{Asadi:2026iot,
    author = "Asadi, Pouya and Batz, Austin and Fox, Patrick J. and Homiller, Samuel D. and Kribs, Graham D.",
    title = "{For Whom the Xenon Recoils: Magnetic Inelastic Dark Baryons}",
    eprint = "2609.09107",
    archivePrefix = "arXiv",
    primaryClass = "hep-ph",
    reportNumber = "PITT-PACC-2614, FERMILAB-PUB-26-0660-T",
    month = "9",
    year = "2026"
}

@article{Yuan:2026djt,
    author = "Yuan, Guan-Wen and Zhang, Bo and Cao, Wen-Yu and Feng, Lei and Yang, Ruizhi",
    title = "{ALP-mediated inelastic dark matter and the LUX-ZEPLIN high-recoil candidate event LZ230616}",
    eprint = "2609.08893",
    archivePrefix = "arXiv",
    primaryClass = "hep-ph",
    month = "9",
    year = "2026"
}

@article{Okada:2026upm,
    author = "Okada, Hiroshi and Shigekami, Yoshihiro and Wu, Jia-Jun",
    title = "{Can a minimal radiative seesaw explain the LZ 248 keV event?}",
    eprint = "2609.13038",
    archivePrefix = "arXiv",
    primaryClass = "hep-ph",
    month = "9",
    year = "2026"
}

@article{Frolovsky:2026tvq,
    author = "Frolovsky, Daniel and Ketov, Sergei V.",
    title = "{Higgsino dark matter in the Starobinsky supergravity with the MSSM in light of the LUX-ZEPLIN event}",
    eprint = "2609.11241",
    archivePrefix = "arXiv",
    primaryClass = "hep-ph",
    reportNumber = "IPMU26-0034",
    month = "9",
    year = "2026"
}

@article{Pospelov:2026ewn,
    author = "Pospelov, Maxim and Ramani, Harikrishnan",
    title = "{Strong Constraints on Higgsino Dark Matter from Solar Capture}",
    eprint = "2609.02775",
    archivePrefix = "arXiv",
    primaryClass = "hep-ph",
    month = "9",
    year = "2026"
}

@article{Rodd:2026tyn,
    author = "Rodd, Nicholas L. and Safdi, Benjamin R. and Slatyer, Tracy R. and Xu, Weishuang Linda",
    title = "{Confronting the Higgsino Interpretation of the LZ Event with the High-Energy Sideband}",
    eprint = "2609.04175",
    archivePrefix = "arXiv",
    primaryClass = "hep-ph",
    month = "9",
    year = "2026"
}

@article{Bose:2026ndd,
    author = "Bose, Debajit and others",
    title = "{Not so good $\nu$s for Higgsino dark matter as LZ excess: stringent limits from Super-Kamiokande and IceCube}",
    eprint = "2609.07807",
    archivePrefix = "arXiv",
    primaryClass = "hep-ph",
    month = "9",
    year = "2026"
}

@article{Nguyen:2026lui,
    author = "Nguyen, Thong T. Q. and Linden, Tim and Hooper, Dan",
    title = "{Solar Neutrino Constraints on Inelastic Dark Matter Scattering in Light of Recent LUX-ZEPLIN Observations}",
    eprint = "2609.11833",
    archivePrefix = "arXiv",
    primaryClass = "hep-ph",
    month = "9",
    year = "2026"
}

@article{DiMauro:2026dqp,
    author = "Di Mauro, Mattia and Shaikh, Halim",
    title = "{Solar Capture Tests of Inelastic Dark Matter after the LZ High-Recoil Event}",
    eprint = "2609.06760",
    archivePrefix = "arXiv",
    primaryClass = "hep-ph",
    month = "9",
    year = "2026"
}

@article{Chatterjee:2026scv,
    author = "Chatterjee, Arindam and Das, Debottam and Pasha, Syed Adil and Pukhov, Alexander and Puri, Rahul",
    title = "{Radiative Corrections to the Direct Detection of Inelastic Scattering of Higgsino-like Neutralino Dark Matter}",
    eprint = "2609.09830",
    archivePrefix = "arXiv",
    primaryClass = "hep-ph",
    month = "9",
    year = "2026"
}

@article{Cheung:2026byg,
    author = "Cheung, Kingman and Kang, Sin Kyu and Kumar, Ranjeet",
    title = "{From LUX-ZEPLIN to Colliders: Probing Higgsino Dark Matter}",
    eprint = "2609.08712",
    archivePrefix = "arXiv",
    primaryClass = "hep-ph",
    month = "9",
    year = "2026"
}

@article{Kotlarski:2026pep,
    author = "Kotlarski, Wojciech and Kowalska, Kamila and Sessolo, Enrico Maria",
    title = "{GUT-induced FCC signatures of the LUX-ZEPLIN event}",
    eprint = "2609.06750",
    archivePrefix = "arXiv",
    primaryClass = "hep-ph",
    month = "9",
    year = "2026"
}

@article{Okada:2016tzi,
    author = "Okada, Nobuchika and Papapietro, Nathan",
    title = "{R-parity Conserving Minimal SUSY $B-L$ Model}",
    eprint = "1603.01769",
    archivePrefix = "arXiv",
    primaryClass = "hep-ph",
    month = "3",
    year = "2016"
}

@article{Oda:2023dqj,
    author = "Oda, Satsuki and Okada, Nobuchika and Papapietro, Nathan and Takahashi, Dai-suke",
    title = "{R-parity Conserving Minimal SUSY U(1)$_{X}$ Model}",
    eprint = "2307.16480",
    archivePrefix = "arXiv",
    primaryClass = "hep-ph",
    month = "7",
    year = "2023"
}

@article{Planck:2018vyg,
    author = "Aghanim, N. and others",
    collaboration = "Planck",
    title = "{Planck 2018 results. VI. Cosmological parameters}",
    eprint = "1807.06209",
    archivePrefix = "arXiv",
    primaryClass = "astro-ph.CO",
    doi = "10.1051/0004-6361/201833910",
    journal = "Astron. Astrophys.",
    volume = "641",
    pages = "A6",
    year = "2020",
    note = "[Erratum: Astron.Astrophys. 652, C4 (2021)]"
}

@article{ATLAS:2019erb,
    author = "Aad, Georges and others",
    collaboration = "ATLAS",
    title = "{Search for high-mass dilepton resonances using 139 fb$^{-1}$ of $pp$ collision data collected at $\sqrt{s}=$13 TeV with the ATLAS detector}",
    eprint = "1903.06248",
    archivePrefix = "arXiv",
    primaryClass = "hep-ex",
    reportNumber = "CERN-EP-2019-030",
    doi = "10.1016/j.physletb.2019.07.016",
    journal = "Phys. Lett. B",
    volume = "796",
    pages = "68--87",
    year = "2019"
}

@article{CMS:2021ctt,
    author = "Sirunyan, Albert M and others",
    collaboration = "CMS",
    title = "{Search for resonant and nonresonant new phenomena in high-mass dilepton final states at $ \sqrt{s} $ = 13 TeV}",
    eprint = "2103.02708",
    archivePrefix = "arXiv",
    primaryClass = "hep-ex",
    reportNumber = "CMS-EXO-19-019, CERN-EP-2021-026",
    doi = "10.1007/JHEP07(2021)208",
    journal = "JHEP",
    volume = "07",
    pages = "208",
    year = "2021"
}

@article{Pumplin:2002vw,
    author = "Pumplin, J. and Stump, D. R. and Huston, J. and Lai, H. L. and Nadolsky, Pavel M. and Tung, W. K.",
    title = "{New generation of parton distributions with uncertainties from global QCD analysis}",
    eprint = "hep-ph/0201195",
    archivePrefix = "arXiv",
    reportNumber = "MSU-HEP-011101",
    doi = "10.1088/1126-6708/2002/07/012",
    journal = "JHEP",
    volume = "07",
    pages = "012",
    year = "2002"
}

\end{document}